\documentstyle[11pt,newpasp]{article}

\input psfig.tex

\begin{document}

\title{The Large Zenith Telescope: Redshifts and Physical Parameters of Faint
Objects from a Principal Component Analysis}

\author{R\'emi A. Cabanac\altaffilmark{1} and Val\'erie de Lapparent}
\affil{Institut d'astrophysique de Paris (CNRS), France}

\altaffiltext{1}{Fellow of the Fonds FCAR, Qu\'ebec, Canada.} 



\section{The Large Zenith Telescope Survey}

The Large Zenith Telescope (LZT) is a 6-m liquid mirror telescope (Hickson 1998;
Borra et al. 1992; Cabanac et al. 1998). It is a transient instrument dedicated
to a survey of faint objects (Table 1). 
The final input of the survey will be a catalog giving the photometric fluxes 
in $\sim40$ narrow-band filters (3700\AA-1$\mu$ range), as well as basic 
morphological information for $\sim10^6$ objects. Here we investigate the 
potential of the Principal Component Analysis (PCA) to extract meaningful 
physical parameters from such a large dataset.

\begin{table}
\caption{The Large Zenith Telescope Survey Characteristics.} \label{tbl-1}
\begin{center}\scriptsize
\begin{tabular}{rl}
\tableline
Latitude&$49^\circ17'17.2''$\\
Median seeing&$0.9''$\\
Diameter of corrected field&$24'$\\
Detector&Thinned 2048 x 2048 CCD\\
Image scale&$0.495''$\\
Integration time&$64.8\,sec$\\
Limiting mag in filter $750\,nm$&24.4\\
Expected \# of stars/galaxies/QSOs&$10^5/10^6/10^4$\\
Area covered in one night (8h)&$0.28^\circ \times 80^\circ$\\
\end{tabular}
\end{center}
\end{table}
\section{Principal Component Analysis and Mock Catalogs}

The application of the PCA to the classification of galaxies is described 
elsewhere (Murtagh \& Heck 1987; Galaz \& de Lapparent 1998 and references 
therein). After a PCA, the spectrophotometric energy distribution (SED) of 
each object can be approximated as $S_{approx}=\alpha_{1}E_{1}+\alpha_{2}E_{2}+
\alpha_{3}E_{3}+\ldots,$ where $S_{approx}$ is the SED, $\alpha_{1}, 
\alpha_{2}, \alpha_{3}\ldots$ are the eigenvalues, $E_{1}, E_{2}, E_{3}\ldots$ 
are the eigenvectors.
We build realistic mock catalogs of LZT SEDs to a limiting magnitude 
$R_{mag} < 23$ for stars (number counts and color distribution from the Bahcall-Soneira model, ARAA 1986), galaxies (Fioc \& Rocca-Volmerange GISSEL and 
Bruzual \& Charlot PEGASE spectra, luminosity functions by Autofib for 
$0<z_{shift}<0.5$, and CFRS luminosity function for $0.5<z_{shift}<2$), 
and QSOs (number counts and counts vs $z_{shift}$ from 2dF QSO Survey).

\section{Results: morphological discrimination and redshift}

(i) Most of the information in the LZT spectral energy distributions is 
contained in the continua. Only strong emission lines are detected 
($W_{line}>50~\AA$).
(ii) The PCA is robust to discriminate different species at a resolution of 
$\sim40$, and at low $S/N$. (iii) Redshifts of galaxies can be derived for
$z_{shift}<1.5$. Strong degeneracies occur at higher redshifts 
between high $z_{shift}$  blue galaxies and local red galaxies because
the H \& K (Ca II) break exits the observed spectral range.

\begin{table}
\caption{Error in PCA-measured redshift ($\sigma_{z}$) vs true redshift ($z_{theor}$)}
\label{tbl-2}
\begin{center}\scriptsize
\begin{tabular}{ccc|ccc|ccc}
$\# objects$&$z_{theor}$&$\sigma_{z}$&$\# objects$&$z_{theor}$&$\sigma_{z}$&
$\# objects$&$z_{theor}$&$\sigma_{z}$\\
\tableline
(star)2963&0.0&0.049&1082&0.7&0.013&373&1.5&0.0103\\
109&0.1&0.134&906&0.9&0.056&234&1.7&0.222\\
567&0.3&0.015&812&1.1&0.084&568&1.9&0.177\\
926&0.5&0.023&568&1.3&0.053&(QSO)41&2.0-2.5&0.0\\
\end{tabular}
\end{center}
\end{table}
%
%
%
\begin{figure}
\psfig{file=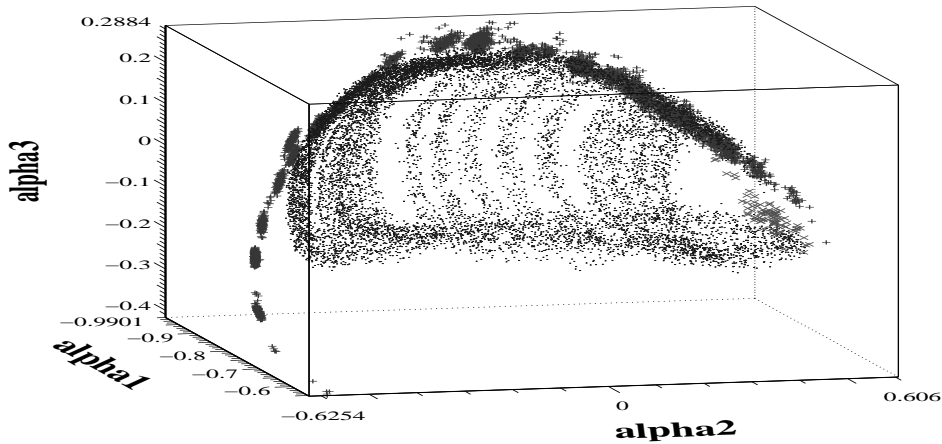}
\caption{3D plot of the first three eigenvalues of a PCA on a mock
catalog ($S/N=3$). Different types of galaxies at different $z_{shift}$
(dots), OBAFGKM stars (+) and QSOs ($\times$) occupy different $loci$.} \label{fig-1}
\end{figure}

\end{document}